\documentclass[prb,aps,twocolumn,amsmath,amsfonts]{revtex4}

\usepackage{graphicx}

\def\Tr{\mbox{Tr}\,}
\def\tr{\mbox{tr}\,}
\def\cst{\mathrm{const}}

\begin{document}

\title{Phase transitions in full counting statistics for periodic pumping}

\author{Dmitri A.~Ivanov}
\affiliation{Institute of Theoretical Physics,
Ecole Polytechnique F\'ed\'erale de Lausanne (EPFL), 
CH-1015 Lausanne, Switzerland}

\author{Alexander G.~Abanov}
\affiliation{Department of Physics and Astronomy,
Stony Brook University,  Stony Brook, NY 11794-3800.}

\date{July 14, 2010}

\begin{abstract}
We discuss the problem of full counting statistics for periodic pumping. 
The probability generating function 
is usually defined on a circle of the ``physical'' values of 
the counting parameter, with its periodicity corresponding to charge 
quantization. The extensive part of the generating function can
either be an analytic function on this circle or have singularities. These 
two cases may be interpreted as different thermodynamic phases
in time domain. We discuss several examples of phase transitions
between these phases for classical and quantum systems. Finally,
we prove a criterion for the ``analytic'' phase in the problem
of a quantum pump for noninteracting fermions.
\end{abstract}

\maketitle

\section{Introduction}

The problem of full counting statistics (FCS) \cite{1993-LevitovLesovik,Nazarov-proc} 
is often considered in a setup
periodic in time.\cite{1997-IvanovLeeLevitov}
In such a formulation, the system depends on
external parameters varying periodically in time, and one is interested
in counting certain quantized events (typically, the transfer of
particles between the leads of a contact). Generally, the 
probabilities of different outcomes $P_n$ are labeled by integer
indices $n$ and can be combined into the probability generating function \cite{1993-LevitovLesovik,2000-BlanterButtiker}
\begin{equation}
	\chi(\lambda) =\sum_{n=-\infty}^{+\infty}P_n e^{i\lambda n}\, .
 \label{gen-function-def}
\end{equation}
For a superposition of statistically independent processes, the
generating function is given by the product of those for each
process. Therefore, for a periodic process extended in time,
this generating function is exponentially extensive in time,
provided the correlations in time decay sufficiently rapidly.
One therefore usually defines the generating function ``per
period''  \cite{1997-IvanovLeeLevitov}  (or ``extensive'' FCS)
\begin{equation}
	\chi_0(\lambda)=\exp\left[\lim_{N\to\infty} 
	\frac{1}{N} \ln \chi(\lambda)\right]\, ,
 \label{chi-extensive}
\end{equation}
where the FCS $\chi(\lambda)$ is collected over $N$ periods.

A vast literature is devoted to FCS 
both in the general setup and for periodic 
processes.\cite{Nazarov-proc}
Probably, the most interesting class of FCS problems are 
those of quantum charge
transfer,\cite{1993-LevitovLesovik,1997-IvanovLeeLevitov}
but there are also discussions of FCS in classical
stochastic processes (see, e.g., Ref.~\onlinecite{2003-BagretsNazarov}) 
and of the relation between classical
and quantum effects in FCS  (see, e.g., Sec.~5 of Ref.~\onlinecite{2000-BlanterButtiker} and Ref.~\onlinecite{2005-SukhorukovBulashenko}).

From this immense body of results, one notices that FCS in time-extensive
problems can be conveniently classified by the analytic properties
of the FCS generating function per period (\ref{chi-extensive}).
Namely, two main phases can be identified: $\ln\chi_0(\lambda)$ 
may either be analytic on the unit circle of real $\lambda$ 
or have a singularity at certain values of $\lambda$. Correspondingly we
distinguish two phases of FCS: {\it analytic} and {\it nonanalytic}.

In the present paper we show that the existence of these two phases
is a very general feature of FCS: phase transitions between them occur 
both in classical stochastic models and in quantum systems.
We illustrate our discussion with several examples: the 
classical weather model and quantum systems of noninteracting fermions.
For quantum examples, we use the results of our recent 
work\cite{2008-AbanovIvanov,2009-AbanovIvanov} to show the stability
of the analytic phase for a wide class of non-interacting fermionic
systems, even at finite temperature. Finally, we discuss 
a possible identification of the two phases in terms of
cumulants and conjecture the general form of the asymptotic 
long-time behavior of the FCS in the nonanalytic phase.

\section{Two phases of FCS in time-periodic systems}

The generating function (\ref{gen-function-def}) is periodic,
$\chi(\lambda)=\chi(\lambda+2\pi)$, and its Fourier components $P_n$ are
non-negative. Furthermore, the probabilities $P_n$ always
obey the normalization condition $\sum_n P_n = 1$, and therefore for
non-negative $P_n$ the series (\ref{gen-function-def}) converges
uniformly on the unit circle $| e^{i\lambda} | =1$. Typically,
$P_n$ decay sufficiently rapidly as functions of $n$, and then
$\chi(\lambda)$ is analytic for real values of $\lambda$.

Out of the above three properties (periodicity, non-negativity of probabilities
and analyticity) only the first one (periodicity) necessarily holds for 
the generating function per period (\ref{chi-extensive})
from its definition. As we shall see below, both the 
non-negativity of the Fourier components  
(``quasiprobabilities'') and the analyticity may be broken for 
$\chi_0(\lambda)$. Note that non-negativity of the quasiprobabilities
is not always related to analyticity (e.g., $\chi_0(\lambda)$ may 
be analytic and still have some negative 
quasiprobabilities).

As we shall see below, most common nonanalityc features 
of $\chi_0(\lambda)$ are discontinuities and kink points. 
In those cases, the quasiprobabilities
decay algebraically in $n$ and oscillate in sign (if the
singularity is located at $\lambda\ne 0)$. On the
other hand, an analytic $\chi_0 (\lambda)$ corresponds 
to an exponential decay of the quasiprobabilities 
(not necessarily non-negative). Thus we may view the
difference between the analytic and nonanalytic behaviors of 
$\chi_0 (\lambda)$ as a distinction between two quantum phases.
As usual for phase transitions, the transition to the nonanalytic
phase appears only in the thermodynamic limit (when one considers the
extensive part of the generating function).

We may also remark in passing that there is nothing surprising in
negative quasiprobabilities in the extensive generating function
$\chi_0(\lambda)$. Indeed, this generating function appears in our
attempt to factorize the full FCS into $N$ independent processes
(\ref{chi-extensive}). Generally, there is no reason to expect that
such decomposition is possible with physical non-negative probabilities, 
since the system has certain correlations between different periods.

\section{Classical example: weather model}

The simplest stochastic process that illustrates the analytic-nonanalytic
phase transition is the so called ``weather model''.\cite{weather-book}
Suppose that the
weather on each day can be of two types: rainy or sunny, and that 
the weather on a given day is chosen randomly with the probabilities 
depending on the weather
on the previous day. Let the probability of a sunny day after a sunny day
be $q_{s}$ and the probability of a rainy day followed by a rainy day be $q_{r}$ (Fig.~1a).
\begin{figure}
	\includegraphics[width=0.25\textwidth]{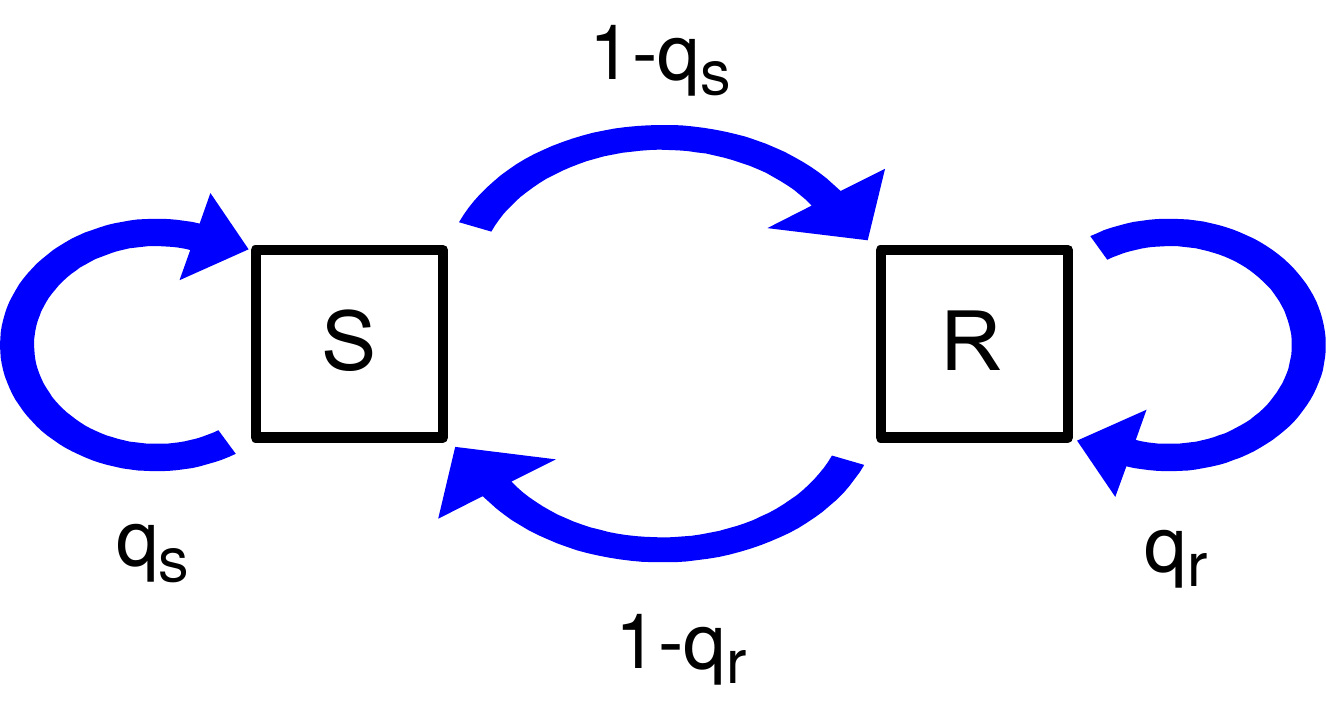}\qquad
	\includegraphics[width=0.12\textwidth]{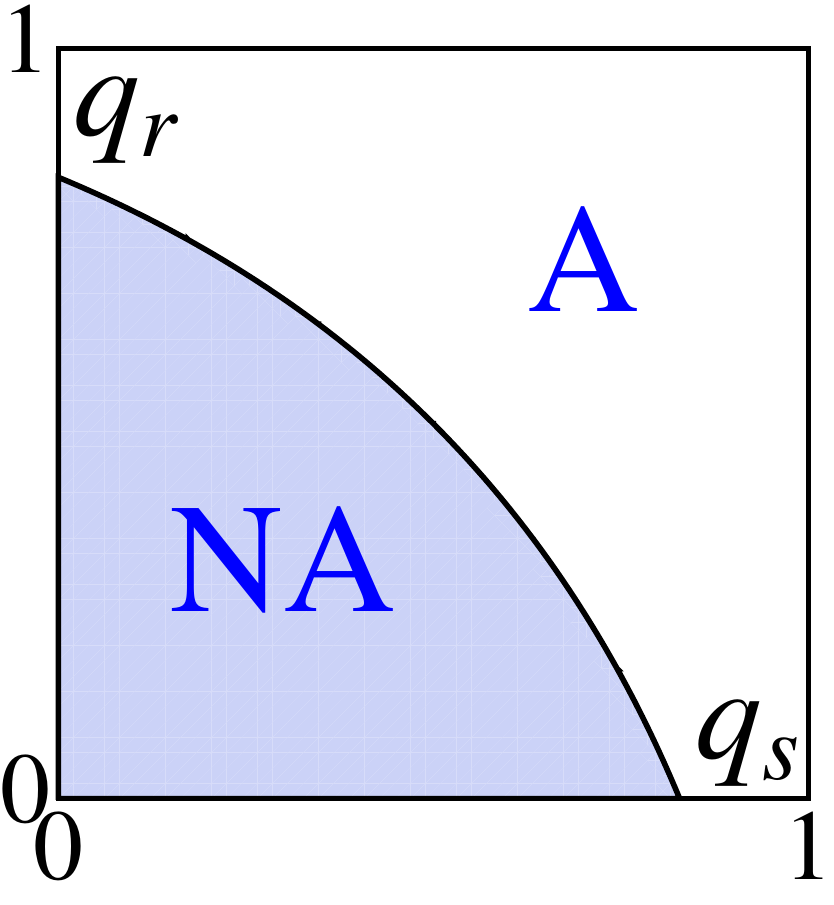}
\caption{{\bf (a)} Weather model. Transitions to sunny (rainy) 
weather S (R) are shown by arrows with corresponding probabilities. {\bf (b)}
Phase diagram of the weather model in the $(q_{s},q_{r})$ parameter space. 
The shaded area corresponds to the nonanalytic phase (NA).}
\label{fig:wmd}
\end{figure}
This defines a stochastic process, in which the FCS for the number of sunny
days can be easily calculated.\cite{2007-vanKampen} The calculation involves two generating
functions $\chi_s(\lambda)$ and $\chi_r(\lambda)$, which correspond to
the conditions that the last day is sunny or rainy, respectively. The
evolution of these generating functions is described by a linear operator
acting in the two-dimensional space. The full generating function is then
given by
\begin{equation}
	\chi(\lambda) = \chi_s(\lambda) + \chi_r(\lambda)
	= C_1 [\omega_1(\lambda)]^N + C_2 [\omega_2(\lambda)]^N\, ,
 \label{chiweather}
\end{equation}
where $N$ is the number of observation days, $\omega_1(\lambda)$ and 
$\omega_2(\lambda)$ are
the two eigenvalues of the evolution operator, and $C_1$ and $C_2$ are some
coefficients. Therefore
\begin{equation}
\chi_0(\lambda) = \max \left(\omega_1(\lambda),\omega_2(\lambda)\right)\, , 
 \label{chi0weather}
\end{equation}
where the eigenvalue with the maximal absolute value is chosen. 
Then two phases are possible: either one
eigenvalue remains leading for all values of $\lambda$ (analytic phase) 
or the leading eigenvalue switches at some value of $\lambda$ 
(nonanalytic phase). From the explicit formula for the eigenvalues 
\begin{equation}
\omega_{1,2}= \frac{q_{r}+q_{s} e^{i\lambda}}{2}\pm \sqrt{\left( 
\frac{q_{r}+q_{s} e^{i\lambda}}{2}\right)^{2}+(1-q_{s}-q_{r}) e^{i\lambda}}
\end{equation}
one can deduce the phase diagram in the $(q_{s},q_{r})$ coordinates, (Fig.~1b) and find that the nonanalyticity occurs at $\lambda =\pi$.

\section{Quantum examples: non-interacting fermions}

As quantum examples illustrating the analytic-nonanalytic 
transition in time-periodic FCS, we consider
systems of non-interacting fermions.
This class of systems has been studied extensively,\cite{Nazarov-proc}
with the main result
given by the so-called Levitov--Lesovik determinant formula
for the characteristic function (\ref{gen-function-def}).
\cite{1993-LevitovLesovik,1993-IvanovLevitov,
1996-LevitovLeeLesovik,1997-IvanovLeeLevitov}
In the periodic case, the same determinant formula can be used to
find the extensive part of the FCS (\ref{chi-extensive}) by
simply imposing periodic boundary conditions
in time.\cite{1997-IvanovLeeLevitov}

For simplicity, we consider here a one-channel quantum contact 
of non-interacting spinless fermions.
The transparency of the contact $g$ is taken to be time-independent,
but we assume a time-dependent voltage $V(t)$ applied to the contact and
an arbitrary temperature $T$. This
system was considered in previous works, \cite{2007-VanevicNazarovBelzig} and we can
use their results to study the analytic-nonanalytic transition
in a periodic setup.
The case of a more general time-dependent scattering matrix \cite{2008-AbanovIvanov,2009-AbanovIvanov} 
will be treated in the next section.

The FCS for non-interacting fermions
can be expressed in terms of the distribution function of 
effective transparencies
$\mu(p)$ for single-particle processes.\cite{2009-AbanovIvanov}
This function determines the jump of
the derivative of $\ln \chi_0(\lambda)$ on the negative real axis 
of the variable $e^{i\lambda}$ [so that $p\in (0,1)$]:
\begin{equation}
	\mu(p)=
	\frac{1}{2\pi i} \partial_{p}\ln\chi(\lambda)
	\Big|^{p-i0}_{p+i0}\, ,
 \quad
 	p=\frac{1}{1-e^{i\lambda}} \, .
 \label{jump-in-derivative2}
\end{equation}

Since, in the non-interacting case, singularities of $\chi_0(\lambda)$ are 
allowed only on the negative real axis of $e^{i\lambda}$ 
(i.e., if we take $\lambda$ to be real, at 
$\lambda=\pi$),\cite{2008-AbanovIvanov,2009-AbanovIvanov}
the analytic and nonanalytic phases correspond to $\mu(p=1/2)=0$ and  
$\mu(p=1/2)\ne 0$, respectively.

\begin{figure}
\includegraphics[width=0.48\textwidth]{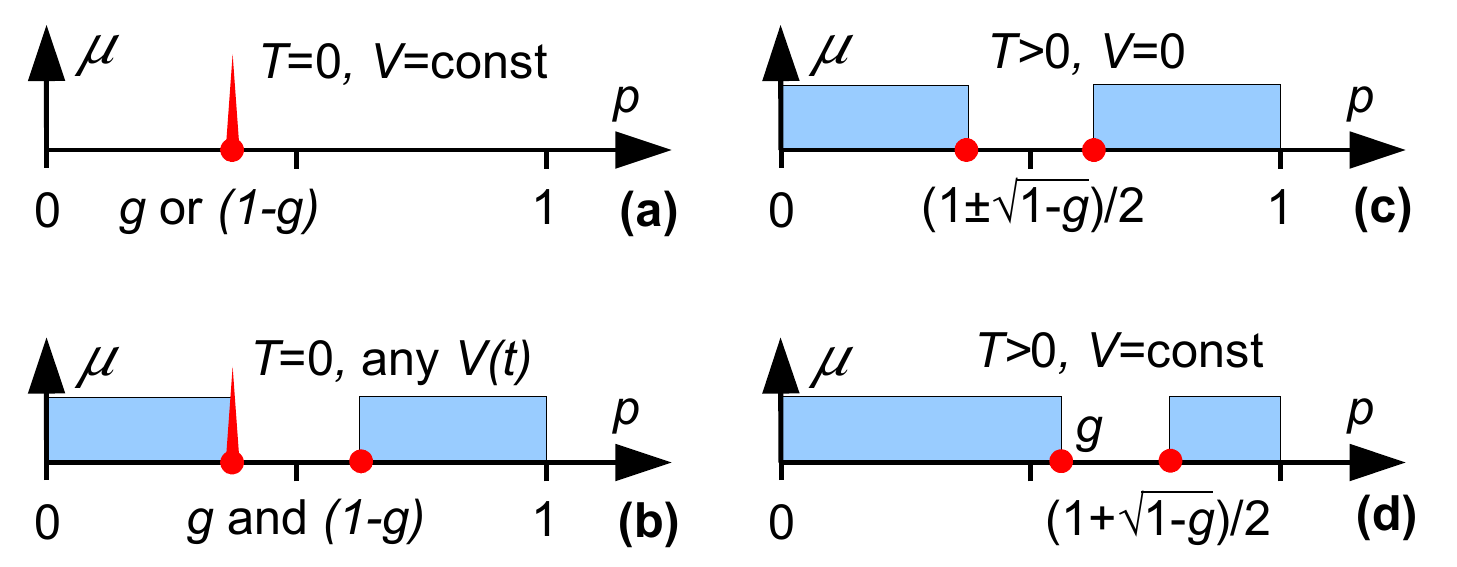}
\caption{Schemes of distributions of effective transparencies $\mu(p)$
for different examples of FCS in a quantum contact. Shaded areas represent
a continuous spectrum $\mu(p)$. Peaks at gap edges schematically
represent delta-function contributions.}
\label{fig:quant}
\end{figure}

From the previous studies of the one-channel contact with a time-independent
transparency $g$, we can sketch the general behavior of the distribution
function $\mu(p)$ in several special cases (Fig.~2):

{\bf (a)} $T=0$, $V(t)=\cst$. In this case $\mu(p)=\alpha \delta(p-g)$ or
$\mu(p)=\alpha \delta(p-[1-g])$, depending on the polarity of $V$ (Fig.~2a). 
The weight $\alpha=V \tau /(2\pi)$, where $\tau$ is the conventional period. This
case obviously belongs to the {\it analytic} phase, unless $g=1/2$.

{\bf (b)} $T=0$, arbitrary $V(t)$. In this case, $\mu(p)$ is given by a 
superposition of the delta function from the previous example and a 
generally continuous
spectrum for $|p-1/2| \ge |g-1/2|$ (gapped around $p=1/2$, except for
$g=1/2$), see Fig.~2b. This form of the spectrum (delta function and a
continuous part) follows from the arguments in 
Refs.~\onlinecite{2007-VanevicNazarovBelzig,2008-AbanovIvanov}). 
In addition, it is
proved there that the continuous part is symmetric with respect to $p=1/2$.
The existence of the gap around $p=1/2$ also follows from the theorem
proved in the next section. Except for $g=1/2$, this case also
belongs to the {\it analytic} phase.

{\bf (c)} $T>0$, $V(t)=0$. In this case, the spectrum $\mu(p)$ is symmetric,
continuous, and gapped around $p=1/2$ (Fig.~2c). The value of the gap 
can be obtained by a direct calculation
\cite{1996-LevitovLeeLesovik,2009-KlichLevitov}: $|p- 1/2| \ge \sqrt{1-g}/2$
(also predicted by the theorem proven in the next section).
This example belongs to the {\it analytic} phase as well.

{\bf (d)} $T>0$, $V(t)=\cst$ (without loss of generality, we assume $V>0$,
otherwise we may simply reflect $p\mapsto 1-p$).
In this case, a calculation shows that the spectrum is
continuous and consists of two regions: $0\le p \le g$ and 
$(1+\sqrt{1-g})/2 \le p \le 1$ (Fig.~2d). This case belongs to
the {\it analytic} phase, if $g<1/2$ and to the {\it nonanalytic}
phase, if $g>1/2$. Note that the two regions of support of $\mu(p)$
overlap if $g \ge 3/4$.

{\bf (e)} $T>0$, arbitrary $V(t)$. In this case, the spectrum $\mu(p)$
is generally continuous and non-symmetric. As we prove in the theorem
in the next section, the {\it analytic} phase is realized for $g<1/2$
(which results in the gap $\left| p-1/2 \right| > 1/2 - g$).
The {\it nonanalytic} phase is generic for $g>1/2$.

\section{Analytic phase for non-interacting
quantum systems}

Let us now consider a slightly more general setup:
a one-channel two-lead contact which can be described by an instantaneous
time-dependent 2$\times$2 scattering matrix $S(t)$. We assume a finite
temperature $T$. The characteristic function 
$\chi(\lambda)$ in this case  is given by \cite{2009-AbanovIvanov} 
\begin{equation}
\chi(\lambda) = 
  \det\left(\left[1+(e^{i\lambda}-1)\tilde{X}\right]e^{-i\lambda Q}\right)\, .
 \label{XlambdaFCS}
\end{equation}
Here the determinant is taken in the single-particle Hilbert space,
$Q=(1+\sigma^{z})/2$ is the charge operator in one of the leads,
and the Hermitian operator $\tilde{X}$ is defined as 
($n_F$ is the Fermi occupation number)
\begin{equation}
  \tilde{X} =(1-n_{F})Q +
  (n_{F})^{1/2} S^{\dagger}Q S (n_{F})^{1/2}.
\label{tX}
\end{equation}
The spectrum of the operator $\tilde{X}$ is real and confined to the
interval $[0,1]$. Its eigenvalues can be interpreted as effective
transparencies of the contact between leads, and its spectral
density $\mu(p) = \tr \delta(p-\tilde{X})$ determines the characteristic
function as
\begin{equation}
	\ln\chi(\lambda)  = -i\lambda \Tr Q 
	+ \int_{0}^{1}dp\,\mu(p)\ln\left[1+(e^{i\lambda}-1)p\right] \, .
 \label{CGFp}
\end{equation}
The same calculation applies to the extensive part of the generating
function $\chi_0(\lambda)$, if one imposes periodic boundary conditions 
in time.\cite{1997-IvanovLeeLevitov}

\begin{figure}
\includegraphics[width=0.25\textwidth]{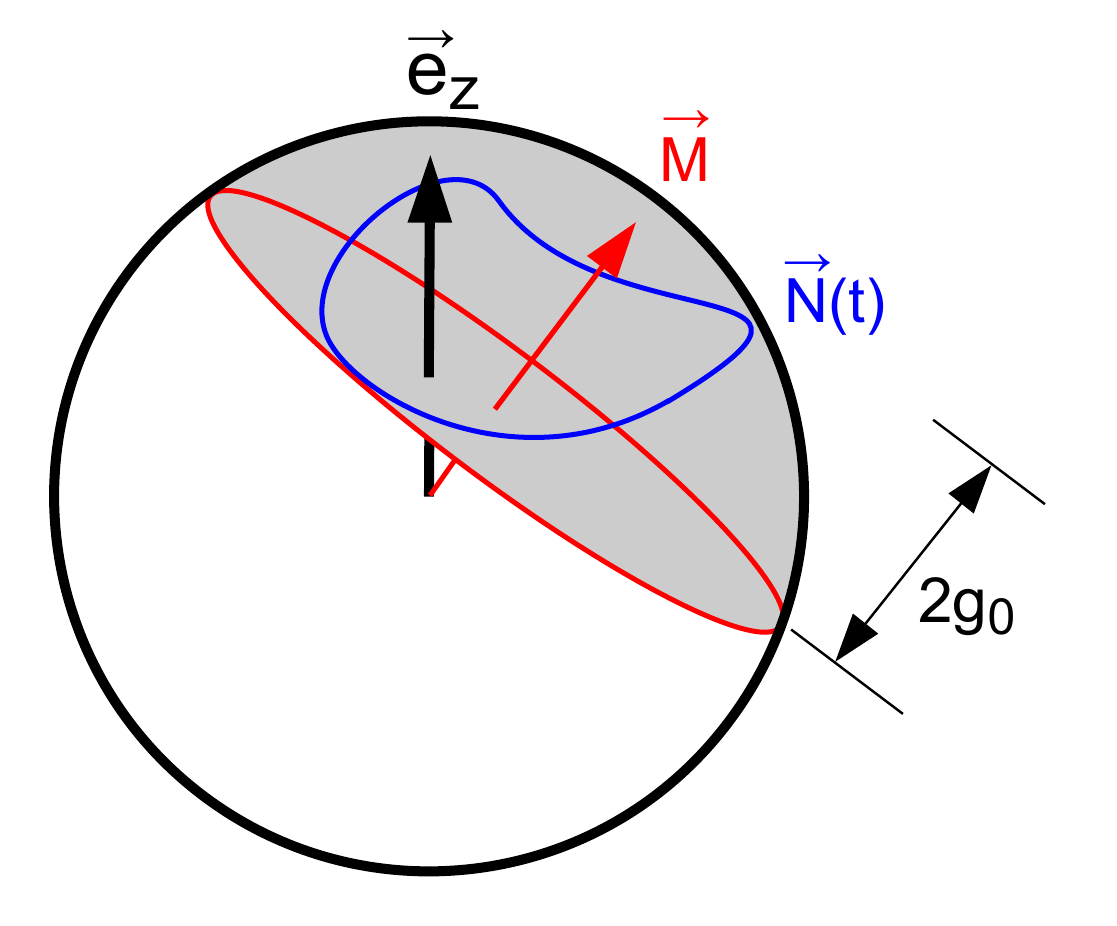}
\caption{Spherical cap of height $2g_0$ covering the trajectory $\vec{N}(t)$
and the north pole $\vec{e}_z$.}
\label{fig:scap}
\end{figure}

As mentioned in the previous section, the analytic or nonanalytic phase
in the periodic setup depends on whether $\mu(p=1/2)$ is zero or non-zero.
The proof of the analytic phase thus amounts to demonstrating a gapped region
around $p=1/2$ for a certain class of time-dependencies 
$S(t)$. Namely, we prove below that if the trajectory of $S(t)$ 
is confined within a certain region (spherical cap), then there is a gap in the spectrum
$\mu(p)$ around $p=1/2$. Consider the matrix 
$\hat{N} = S^{\dagger}\sigma^{z}S$ or, equivalently, the vector 
$\vec{N} = (1/2)\Tr \vec{\sigma}\hat{N}$ 
(it is a time-dependent vector on the unit sphere \cite{2001-MakhlinMirlin}). We suppose
that this vector, at all times, together with the north pole 
$\vec{N}=\vec{e}_z$, fit in a certain spherical cap of height $2g_0<1$ (see Fig.~\ref{fig:scap}).
Then we can prove that $\mu(p)=0$ within the window $|p-1/2|< 1/2-g_0$.

Indeed, the above assumption may be expressed mathematically as
the existence of a constant unit vector $\vec{M}$ such that 
\begin{equation}
\vec{M} \vec{N}(t) \geq 1-2g_{0} \quad  \text{for all } t
\text{ and} \quad
\vec{M} \vec{e}_z\geq 1-2g_{0}\, .
\label{hemisphere}
\end{equation}
We further denote
$\hat{M}=\vec{M}\vec{\sigma}$ and assume that $n_F$ is a scalar
in the lead space (the same temperature in both leads), in particular
$[n_F,\hat{M}]=[n_F,\sigma^z]=0$. Then, after a simple algebra,
\begin{multline}
	\left(2\tilde{X}-1\right)^{2}
	\geq - (\alpha\hat{M})^{2} +\left\{2\tilde{X}-1,\alpha \hat{M}\right\} 
 \\
	= -\alpha^{2} +  \alpha \left(\{ \hat{M}, \sigma^z \} (1-n_F) 
	+ n_{F}^{1/2}\{ \hat{M}, \hat{N} \} n_{F}^{1/2} \right)\,.
\end{multline}
This inequality holds for any real $\alpha$. In particular, we
can choose $\alpha=1-2g_{0}\geq 0$. If we now use the fact 
that the eigenvalues of $n_{F}$ are all located between $0$ and $1$, 
we arrive at the desired inequality
\begin{equation}
	\left(\tilde{X}-1/2\right)^{2} \geq \Big(1/2-g_0\Big)^{2}\, . 
  \label{fTgap}
\end{equation}
This result implies a gap in $\mu(p)$ of the size $1-2 g_0$ centered
around $p=1/2$, i.e., the {\it analytic} phase.

The class of trajectories for which our theorem applies is quite wide.
One particular case are those trajectories, for which the
{\it instantaneous} transparencies $g(t)=(1-\vec{N}(t) \vec{e}_z)/2$
are bounded from above by a certain maximal transparency $g_0<1/2$.
In this case, one may safely choose $\hat{M}=\sigma^z$ and prove the
analytic phase.

Interestingly, in the example (c) of the previous section ($T>0$ and 
time-independent $S$), it is more advantageous to tilt the vector $\vec{M}$
so that it bisects the angle between $\vec{e}_z$ and $\vec{N}$: this
explains the gap $\sqrt{1-g}$ in the spectrum $\mu(p)$ in this 
example.\cite{1996-LevitovLeeLesovik,2009-KlichLevitov}

Another interesting particular case is $T=0$. In this case,
$\mu(p)$ is invariant with respect to global rotations of the trajectory
$\vec{N}(t)$ (see Ref.~\onlinecite{2001-MakhlinMirlin}). This implies that
the condition $\vec{M} \vec{e}_z\geq 1-2g_{0}$ does not need
to be taken into account: it suffices to cover only
the trajectory $\vec{N}(t)$ (and not the north pole) 
by any spherical cap smaller than hemisphere,
in order to prove the analytic phase.
In particular, the transparencies $g$ and $1-g$ produce the same gaps
in examples 1 and 2 of the previous section.

Note that the above theorem may also be extended to the case of different
temperatures of both leads. In this case, we are restricted to 
$\hat{M}=\sigma^z$ (the condition $[n_F,\sigma^z]=0$ is still assumed
as the absence of initial entanglement of the leads\cite{2009-AbanovIvanov}).

Finally, the theorem also applies to the multichannel problem, with
the only modification that the conditions (\ref{hemisphere}) should
be replaced by $(1/2)\{ \hat{M}, \hat{N}(t) \} \geq 1-2g_{0}$
and  $(1/2) \{ \hat{M}, \sigma^z \} \geq 1-2g_{0}$, in terms
of all eigenvalues.

\section{Discussion of results}

We have identified two phases of FCS, analytic and nonanalytic,
in terms of the analyticity of the generating
function per period (\ref{chi-extensive}) at real values of the
counting parameter $\lambda$.
Although singularities of $\chi_0(\lambda)$ in all 
examples considered in the present paper appeared 
at $\lambda=\pi$, they may generally occur at any values of $\lambda$. 
We suggest that the appearance of those singularities 
may be the most common scenario
of the analytic--nonanalytic phase transition and ask the
natural question: what are the fingerprints of such a transition?

The answer to this question may be deduced from analyzing 
classical models (e.g., the weather model described above).
Let us define the {\it staggered average}: $\langle A(n) \rangle_\pi
= \langle (-1)^n A(n) \rangle / \langle (-1)^n \rangle$ for any function
of the counted events $A(n)$.\footnote{Here, 
$\langle A(n)\rangle\equiv \sum_{n}P_{n}A(n)$.} 
Then the {\it staggered cumulant}
\begin{equation}
	\langle\langle n^2 \rangle\rangle_\pi = \langle n^2 \rangle_\pi -
	\langle n \rangle_\pi^2 =
	(-i\partial_{\lambda})^2\ln \chi(\lambda)|_{\lambda=\pi} 
\end{equation}
has different asymptotics as a function of the
number of periods $N$ in the two phases. In the {\it analytic} phase,
$\langle\langle n^2 \rangle\rangle_\pi \propto N$, while in the
 {\it nonanalytic} phase, two eigenvalues $\omega_{1,2}(\lambda)$ contribute,
and one finds $\langle\langle n^2 \rangle\rangle_\pi \propto N^2$.

In quantum noninteracting systems, the same distinction between the two phases
can me made if the asymptotic behavior of the generating function 
$\chi(\lambda)$
over $N$ periods is given in the {\it nonanalytic} phase by 
\begin{equation}
	\chi(\lambda) = C_1  [\chi_0(\lambda-0)]^N
	+ C_2  [\chi_0(\lambda+0)]^N \, ,
\label{Fisher-Hartwig}
\end{equation}
where $\lambda\pm 0$ refers to taking the analytic branches below and
above the cut at $\lambda=\pi$. The coefficients $C_{1,2}$ are
non-extensive (sub-exponential in $N$) and generally depend on
$\lambda$. 
In some situations (e.g., for transmission of non-interacting fermions through an open channel or for counting one-dimensional fermions 
on a line segment at zero temperature), these coefficients are
known to have an algebraic time dependence, 
$C_{1,2} \propto N^{\gamma(\lambda\mp 0)}$, and thus contribute
logarithmically to cumulants. \cite{1994-LevitovLesovik}
In fact, in these simple examples, the expression (\ref{Fisher-Hartwig}) 
can be rigorously proven using the
extended Fisher-Hartwig conjecture 
(theorem),\cite{2010-DeiftItsKrasovsky,2008-Hassler}
as the FCS in those cases is given exactly by a Toeplitz determinant.

We may further conjecture that the
expression (\ref{Fisher-Hartwig}), 
with $C_{1,2}$ algebraically depending on $N$,
should remain valid for a larger
class of models: for example, for all noninteracting fermionic 
systems,\cite{2009-AbanovIvanov}
where the FCS problem can be reduced to
the determinant of a block-Toeplitz matrix. It may even be possible
that this result is applicable for interacting systems. These interesting
problems remain for future study.

Note that a phase transition in FCS in a particular 
interacting system has been recently discussed in 
Ref.~\onlinecite{2010-KarzigVonOppen}, with the difference that their
phase transition reveals itself in the ordinary (non-staggered) cumulants
and is likely connected to 
a nonanalyticity in $\chi_0(\lambda)$ at $\lambda=0$ 
(cf. $\lambda=\pi$ in our examples). 
One can also invent other examples of phase transitions
with singularities appearing at various values of $\lambda$. For example,
in a superconducting system with the charge transfer quantized
in pairs of electrons, a singularity 
would appear at $\lambda=\pi/2$, 
(see example V.A in Ref.~\onlinecite{2009-AbanovIvanov}).

\section{Acknowledgments}

We have  benefited from discussions
with T.~Giamarchi, L.~Levitov, A.~Mirlin,  and E.~Sukhorukov. 
A.G.A. is grateful to ITP, EPFL for hospitality.  
A.G.A. was supported by the NSF under the grant DMR-0906866.



\end{document}